\documentclass[aps,prl,showpacs,twocolumn,floats]{revtex4-2}
\usepackage{amssymb}
\usepackage{amsbsy}
\usepackage{amsmath}
\usepackage{graphicx}
\usepackage{amsmath}
\usepackage{times}
\usepackage{color}
\usepackage{subfigure}
\usepackage{setspace}
\usepackage{bm}

\begin{document}

\title{Exact Floquet flat band and heating suppression via two-rate drive protocols}

\author{Tista Banerjee$^1$, Sayan Choudhury$^2$, and K. Sengupta$^1$}
\affiliation{$^1$School of Physical Sciences, Indian Association for the Cultivation of
Science, Jadavpur, Kolkata-700032, India. \\ $^2$Harish Chandra Research Institute, A CI of Homi Bhabha National Institute,
Chhatnag Road, Jhunsi, Prayagraj, Uttar Pradesh 211019, India. }

\date{\today}

\begin{abstract}

We demonstrate the existence of exact Floquet flat bands implying strong violation of the eigenstate thermalization hypothesis 
in a large class of closed quantum many-body systems in the presence of a two-rate drive characterized by frequencies 
$\Omega_1$ and $\Omega_2=\nu \Omega_1$. We provide the exact analytic condition for this phenomenon 
to occur for a generic protocol; in particular, $\nu=(2p+1)$, where $p$ is an integer, leads to such flat bands 
for both square-pulse and cosine drive protocols for arbitrary $\Omega_1$. In the vicinity of these points, 
heating is suppressed up to very long timescales in such driven systems, leading to a prethermal regime; we demonstrate this by 
exact numerical studies of distribution and bandwidth of the Floquet eigenstates, spectral form factor, entanglement entropy, and correlation functions 
of an experimentally realizable finite driven Rydberg chain. The corresponding micromotion 
exhibits coherent reversal of excitations reminiscent of echoes. Our analysis constitutes a yet unexplored 
mechanism for heating suppression in driven closed quantum systems.  

\end{abstract}


\maketitle

Eigenstate thermalization hypothesis (ETH) predicts eventual 
thermalization for dynamics of quantum systems with initial states which are far from equilibrium \cite{rev1,rev2,rev3,rev4}.
For a periodically driven system governed by its Floquet Hamiltonian, such a steady state 
is characterized by infinite temperature~\cite{dalessio1}; the corresponding Floquet eigenstates  
have uniform probability distribution over the Floquet Brillouin zone (FBZ) leading to 
a bandwidth equal to the drive frequency. 

The violation of ETH stems from the loss of ergodicity arising out of integrability~\cite{rev1}, 
presence of strong disorder~\cite{mblrev}, constraint induced Hilbert space fragmentation ~\cite{fragrev}, or 
presence of quantum scars~\cite{scarref1,scarref2}. The Floquet analogue of such deviations have also been studied~\cite{flmbl1,flscar1,flfrag1};
they typically occur in regimes of high drive amplitudes or frequencies and persist up to long prethermal timescales. However such effects are usually
absent in the low or intermediate drive frequency regime where rapid thermalization occurs over a few drive cycles. 

Such rapid heating in driven quantum systems is detrimental to the coherent control of quantum devices for quantum state preparation and qubit operations; for example, dissipation is often used to counter heating effects in quantum state preparation~\cite{adolrev1,qdevice}. This has 
led to several suggestions for different drive protocols which minimizes heating such as 
counter-diabatic driving~\cite{berry1,adolfo1,polkov1} or the method of optimal control~\cite{opcollit}. However, to the best of our knowledge, 
neither of these methods have been successfully applied for reduction of heating in a generic driven non-integrable many-body system over long time scales.

In this work, we show that a two-rate periodic drive protocol~\cite{adolrev1,jsau1,udiv1,torres1,jaf1,twoexp} characterized by frequencies $\Omega_1$ and 
$\Omega_2= \nu \Omega_1$ ($\nu \in Z$) can lead to reduction of heating in a large class of non-integrable quantum many-body systems. We provide a generic condition  where such a drive leads to Floquet flat bands. For example, both square-pulse and cosine drive protocols with arbitrary $\Omega_1$ and $\nu=(2p+1)$ ($p \in Z$) leads to realization of these flat bands. In contrast to their counterparts studied in non-interacting
driven models \cite{flflat1,flflat2,flflatexp}, these Floquet flat bands are {\it exact eigenstates of the Floquet Hamiltonian} and lead to a strong violation of Floquet ETH. We study such driven systems around the parameter regime where flat bands occur; we find strong suppression of heating over large prethermal timescales in the intermediate and large drive frequencies. The micromotion of such Floquet systems exhibit coherent reversal of excitations which is reminiscent of many-body echo protocols \cite{echoref1,echoref2,echoref3}. These properties lead to qualitative distinction of such driven systems from their single-rate counterparts. We provide a concrete example of this phenomenon using exact diagonalization (ED) on an experimentally realizable Rydberg atom chain via study of its correlation functions, entanglement entropy, spectral form factor (SFF), and the distribution and bandwidth of its Floquet eigenstates. 

\begin{figure}
\rotatebox{0}{\includegraphics*[width= 0.49 \linewidth]{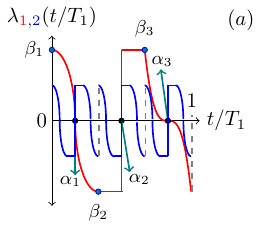}}
\rotatebox{0}{\includegraphics*[width= 0.49 \linewidth]{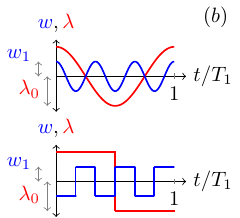}}
\caption{(a) Schematic representation of the two-rate drive protocol with
$\lambda_{1}(t/T_1)$ (red line) and $\lambda_2(t/T_1)$ (blue line) for which $H(t)= \lambda_1(t) \hat O_1 + \lambda_2(t) \hat O_2$ hosts
Floquet flat bands. (b) Specific examples of the cosine (above, Eq.\ \ref{cosp}) 
and square-pulse (below, Eq.\ \ref{sqp}) protocols  with $\nu=3$, $\lambda_0=2w_1=1$ and 
$w_0=0$. \label{fig1}}
\end{figure}

{\it Exact Floquet Flat Bands}: We consider a generic Hamiltonian driven by a two-rate protocol $H(t)= \sum_{i=1,2} \lambda_i(t) \hat O_i$.
Here $\lambda_{i}(t)$ are periodic functions of time with time period $T_i= 2\pi/\Omega_i$ with $i=1,2$, $\Omega_2= \nu \Omega_1$, and $\hat O_i$ are generic many-body  operators with $[\hat O_1,\hat O_2] \ne 0$. For integer $\nu$, the drive has a time period of $T_1$. The protocol chosen is schematically illustrated in Fig.\ \ref{fig1}(a). Importantly, these protocols exhibit turning points at $t_j= \beta_j T_1$ leading to $\lambda_{1,2}(\alpha_j T_1)=0$ for $\alpha_j= (\beta_{j+1} +\beta_j)/2$ as shown; this implies $H(\alpha_j T_1)=0$.  The two-rate generalization of several well-studied one-rate protocols meet this criteria. For instance, the square pulse protocol 
\begin{eqnarray} 
\lambda_1(t) &=&  +(-) \lambda_0 \quad {\rm for} \,\, t \le (>) T_1/2 \nonumber \\
\lambda_2(t) &=& w_0 -[+]w_1 \,\, {\rm for}\,\,  \frac{(m-1)[m]T_1}{2 \nu} \le t < \frac{m[(m+1)]T_1}{2\nu} \nonumber \\
\label{sqp}
\end{eqnarray} 
with $w_0=0$, $ m=1,3, \ldots 2\nu-1$, and $\nu \in Z$ represents such a drive with $\beta_1=0$ and $\beta_2=1$ (bottom panel of Fig.\ \ref{fig1}(b)). Similarly, the cosine protocol (top panel of Fig.\ \ref{fig1}(b)) 
\begin{eqnarray} 
\lambda_1(t) &=& \lambda_0 \cos \Omega_1 t, \quad  
\lambda_2(t) = w_0 + w_1 \cos \nu \Omega_1 t \label{cosp}
\end{eqnarray}
for $\nu=2p+1$ ($p \in Z$) and $w_0=0$ is another example which corresponds to $\beta_1=0$, $\beta_2=1/2$ and $\beta_3=1$. For both protocols, $w_0$ allows one to tune proximity to the flat band limit at $w_0=0$.

As shown in Fig.\ \ref{fig1}, for $w_0=0$, $\lambda_i( \alpha_j T_1 + t_0) = - \lambda_i( \alpha_j T_1 - t_0)$ for all $t_0 \le (\beta_{j+1}-\beta_j)T_1/2$, so that
\begin{eqnarray} 
H( \alpha_j T_1 + t_0) = -H( \alpha_j T_1 - t_0)  \label{hamprop}
\end{eqnarray}
for all $\alpha_j$ and $t_0$. The evolution operator for such a drive protocol can be written as $U(T_1,0) = {\mathcal T}_t \exp[-i \int^t dt' H(t')/\hbar]$, where ${\mathcal T}_t$ indicates time ordering. We write $U(T_1,0)$
using Suzuki-Trotter decomposition and with time steps $\Delta t= T_1/(N_0+1)$ so that  
\begin{eqnarray}
U(T_1,0) = \prod_{k=0\ldots N_0} e^{- i H(t_k) \Delta t/\hbar} = \prod_k U (t_{k+1},t_k) = \prod_k U_k. \nonumber
\end{eqnarray} 
Next, we group $U_k$s between any two turning points $\beta_j$ and $\beta_{j+1}$ and reorganize the Trotter product for $U(T_1,0)$ to write
\begin{widetext}
\begin{eqnarray} 
U(T_1,0) &=& \prod_{j=j_{\rm max}}^1 U(\beta_{j+1} T_1, \beta_{j+1}T_1 -\Delta t) U(\beta_{j+1} T_1-\Delta t, \beta_{j+1}T_1 -2\Delta t)  \ldots   U(\alpha_j T_1 +2\Delta t, \alpha_j T_1 +\Delta t) \nonumber\\
&& \times  U(\alpha_j T_1, \alpha_j T_1 -\Delta t)  \ldots   U(\beta_j T_1+2\Delta t, \beta_j T_1 +\Delta t) U(\beta_j T_1 +\Delta t, \beta_j)\label{ueq1}
\end{eqnarray} 
\end{widetext}
where $j_{\rm max}$ corresponds to the last turning point, $\beta_{j_{\rm max}+1}=1$, and we have used the fact that $U(\alpha_j T_1 +\Delta t, \alpha_j T_1) = I$ since $H(\alpha_j T_1)=0$. Using Eq.\ \ref{hamprop}, we find that the terms in the 
first line Eq.\ \ref{ueq1} is the {\it exact Hermitian conjugate} of those in the second line. This leads to $U(T_1,0)=I$ for such protocols. Since for any periodic drive $U(T_1,0)= \exp[-i H_F T_1/\hbar]$, this yields $E_n^{F}(T_1) =0$ for all Floquet quasi-energies leading to an {\it exact flat band} for any $\Omega_1$. We note that such flat bands have no analogue for single rate drive protocols. They indicate complete localization of all Floquet eigenstates leading to a strong violation of the ETH which predicts uniform distribution of Floquet eigenstates within the FBZ. 

\begin{figure}
\rotatebox{0}{\includegraphics*[width= 0.48 \linewidth]{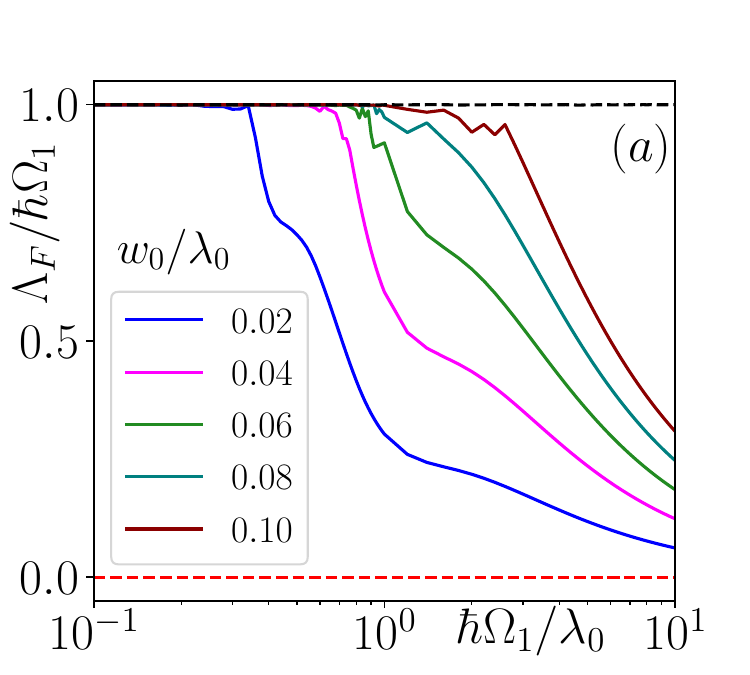}}
\rotatebox{0}{\includegraphics*[width= 0.48 \linewidth]{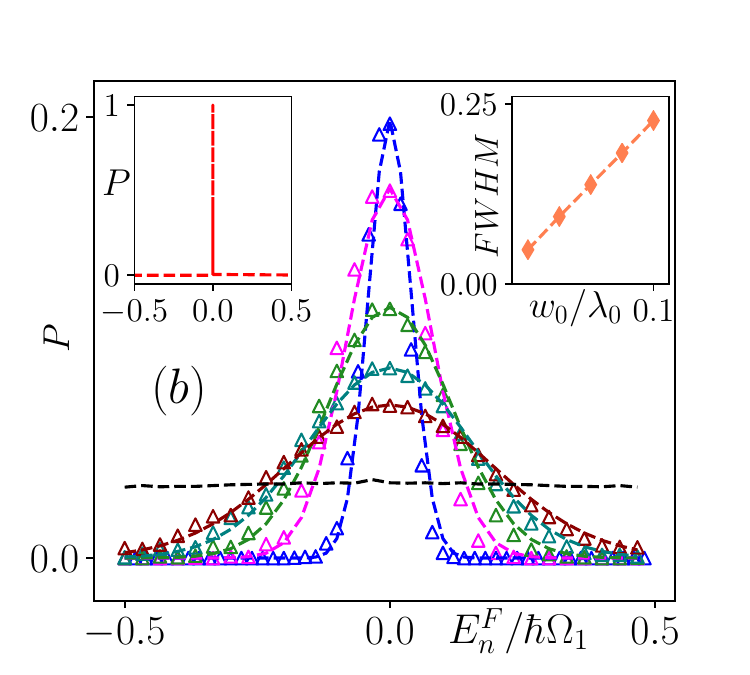}}
\rotatebox{0}{\includegraphics*[width= 0.48 \linewidth]{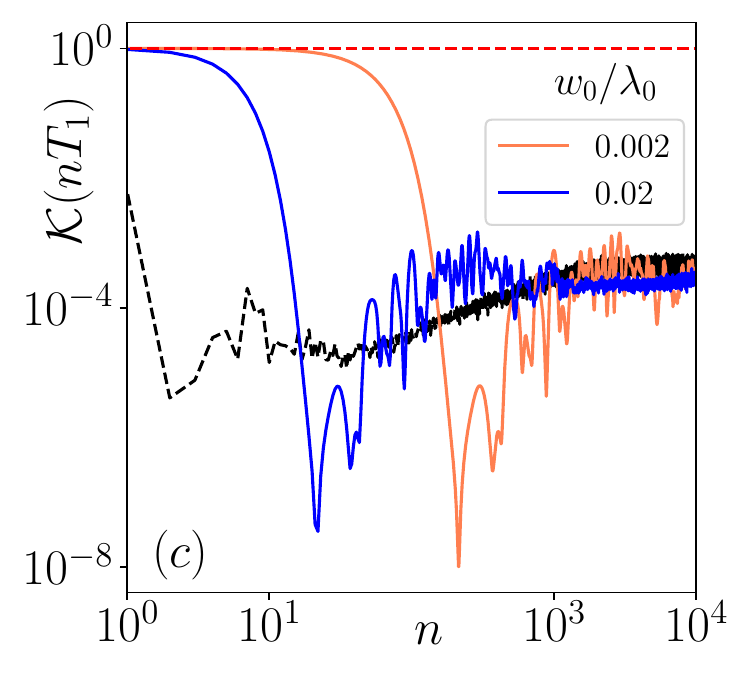}}
\rotatebox{0}{\includegraphics*[width= 0.48 \linewidth]{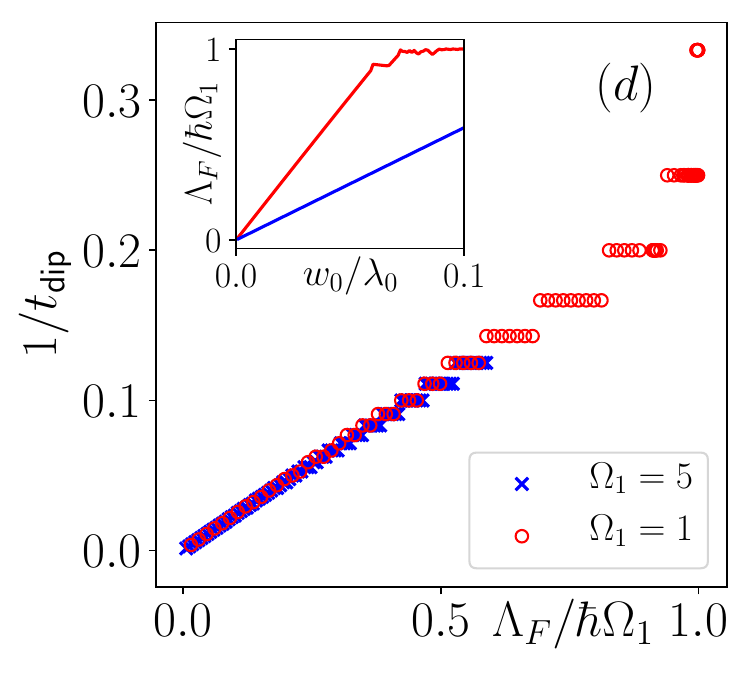}}
\caption{(a) Plot of (a) $\Lambda_F/(\hbar \Omega_1)$ as a function of $\hbar \Omega_1/\lambda_0$ and (b) the distribution $P$ of Floquet eigenvalues $E_n^F$ for $\hbar \Omega_1/\lambda_0=1$ in the first FBZ for a Rydberg chain with $L=26$ in the presence of a square-pulse two-rate protocol with $\nu=3$, $w_1/\lambda_0=1$ and for several representative values of $w_0/\lambda_0$.  The left inset of (b) shows $P$ for $w_0=0$ 
while the right inset shows the FWHM of $P$ as a function of $w_0/\lambda_0$. (c) Plot of ${\mathcal K}(nT_1)$ as a 
function of $n$ for $L=16$, $\hbar \Omega_1/\lambda_0=1$, and several $w_0/\lambda_0$ showing the shift of the dip time. 
(d) Plot of $1/t_{\rm dip}$ as a function of $\Lambda_F/(\hbar \Omega_1)$ for $\hbar \Omega_1/\lambda_0=1$ (red dots) and $5$ (blue cross). The inset shows variation $\Lambda_F/(\hbar \Omega_1)$ with $w_0/\lambda_0$ at these frequencies. For all plots the black (red) dashed line 
corresponds to $w_1/\lambda_0=0(1)$ and $w_0/\lambda_0=1(0)$. See text for details. \label{fig2}}
\end{figure}

{\it Specific Model:} We now consider a specific non-integrable spin-model given by \cite{rydref1,rydref2}
\begin{eqnarray} 
H_R &=& \lambda_1(t) \sum_j \sigma_j^z + \lambda_2(t) \sum_j \tilde \sigma_j^x \label{rydmodel}
\end{eqnarray}  
where $\sigma_j^{x,z}$ denotes Pauli matrices on site $j$, and $\tilde \sigma_j^x = P_{j-1} \sigma_j^x P_{j+1}$, where $P_j= (1-\sigma_j^z)/2$.
Such a model allows spin-flip on site $j$ only if the neighboring sites are in the spin-down state. It is well-known that this model describes the dynamics of a Rydberg atom array of length $L$ in the so-called PXP limit \cite{rydref1,rydref2}, with the Rydberg excitation density given by $\hat n_j = (1+\sigma_j^z)/2$ and the detuning  $\Delta \equiv \lambda_1$. Here we shall study the properties of this model in the presence of a two-rate drive protocols (Eqs.\ \ref{sqp} and \ref{cosp}) in a non-perturbative regime ($\lambda_0 = w_1$) where a single-rate protocol exhibits rapid heating. The corresponding perturbative regime ($\lambda_0 \gg \hbar \Omega_1, w_1, w_0$) is studied by analytical techniques in Ref.\ \cite{supp1}.  

{\it Floquet Eigenstates}: We use ED to obtain exact Floquet eigenvalues and eigenstates of $H_R$ for the square-pulse protocol (Eq.\ \ref{sqp}); the corresponding details and results for cosine protocol (Eq.\ \ref{cosp}) are presented in Ref.\ \cite{supp1}. We first study the normalized Floquet bandwidth $\Lambda_F/(\hbar \Omega_1)$ and the distribution of the Floquet eigenstates $P(E_n^F/\hbar \Omega_1) \equiv P$ over the first FBZ ($-1/2 \le E_F^n/(\hbar \Omega_1) \le 1/2$) for several $w_0$ and the zero total momentum ($K_0=0$) and even parity ($P_0=1$) sector. The results are shown in Fig.\ \ref{fig2} (a) and (b) for several representative $w_0/\lambda_0$ and $w_1/\lambda_0=1$. In this regime, a single-rate drive protocol with frequency $\Omega_1$ and $w_1=0$ exhibits $\Lambda_F/(\hbar \Omega_1)\simeq 1$ for all $\hbar \Omega_1/\lambda_0 \le 10$ and $P \sim {\rm constant}$ for $\hbar \Omega_1 =\lambda_0$ as shown by the black dotted lines in Figs.\ \ref{fig2}(a) and (b); these results are consistent with the prediction of ETH. 

In contrast, for the two-rate protocol with $w_1/\lambda_0=1$, we find a perfect flat band ($\Lambda_F=0$) (for all $\Omega_1$ and $w_0=0$; see red dotted line in Fig.\ \ref{fig2}(a)) for which $P \sim \delta(E_n^F)$ (left inset of Fig.\ \ref{fig2}(b)). For finite $w_0/\lambda_0$, $\Lambda_F/\hbar\Omega_1 <1$ for a wide range of $\Omega_1$ showing violation of ETH in a finite chain near the flat band limit. The plot of $P$ shown for representative $w_0/\lambda_0$ in Fig.\ \ref{fig2}(b) for $\hbar \Omega_1/\lambda_0=1$ indicates that it is approximately Gaussian around $E_n^F=0$ indicating a deviation from the ETH prediction. The full width at half maxima (FWHM) of $P$ increases linearly with $w_0/\lambda_0$ for $w_0 \ll \hbar \Omega_1$ (right inset of Fig.\ \ref{fig2}(b)); however, for the system sizes and frequency ranges studied here, it never reaches close to the ETH predicted uniform distribution. 

Next, we analyze the SFF, a key indicator of quantum chaos for driven systems, which is given by
\begin{eqnarray} 
{\mathcal K}(nT_1) &=& \frac{1}{{\mathcal D^2}} \sum_{p,q=1}^{{\mathcal D}} e^{i(E_p^F-E_q^F)nT_1/\hbar} \label{sffeq} 
\end{eqnarray}
where ${\mathcal D}$ is the Hilbert space dimension. For a driven ergodic system ${\mathcal K}$ displays a characteristic dip-ramp-plateau structure \cite{chalker1,huse1,prosen1}. 
The occurrence of the first dip provides an estimate of the bandwidth, the position of the ramp indicates thermalization time, while the final plateau is expected to occur around the Heisenberg time $t_H$ \cite{sffref1}.  Fig.\ \ref{fig2}(c) shows a plot of ${\mathcal K}$ as a function of $n$ obtained after averaging over $n_0=100$ values of $w_0$ obtained from an uniform distribution with $\delta w_0= 0.1 w_0$. This averaging procedure is known to reduce oscillations in ${\mathcal K}$ and is not otherwise central to our main results \cite{chalker1,huse1,sffref1}. 

Fig.\ \ref{fig2}(c) shows that for $\hbar \Omega_1 =w_1=\lambda_0$, the initial dip never occurs for $w_0=0$ (red dashed line) while for the single rate drive protocol it occurs at $n=n_d \simeq 1$. An estimate of the dip time can be obtained by noting that at short times ${\mathcal K}(nT_1) \simeq 1- (nT_1)^2 \sum_{p,q} (E_p^F-E_q^F)^2/(2\hbar^2 {\mathcal D}^2)$. We convert the sum over eigenstates to an integral over energy gaps $\epsilon$ with a corresponding density of states $\rho = \rho_0 \Lambda_F^{-1} f(\epsilon/\Lambda_F)$ (where $\rho_0 \sim {\mathcal D}$ is determined by $\int_{-\Lambda_F/2}^{\Lambda_F/2} \rho d \epsilon = {\mathcal D} = \sum_m $) to obtain 
\begin{eqnarray}
{\mathcal K}(nT_1) &\simeq&  1- \frac{(n T_1 \Lambda_F)^2\rho_0 }{2 {\mathcal D} \hbar^2} 
\int_{-1/2}^{1/2} dx f(x) x^2 \nonumber\\ 
&=& 1-c_0 (n T_1 \Lambda_F/\hbar)^2 \label{keq}
\end{eqnarray} 
where $x=\epsilon/\Lambda_F$ and $c_0$ is a non-universal constant independent of ${\mathcal D}$ and $\Lambda_F$. Eq.\ \ref{keq} therefore estimates a dip time $t_{\rm dip} = n_d T_1$, where $n_d \sim {\rm Int}[\hbar /(T_1 \Lambda_F)]$ and ${\rm Int}$ denotes the nearest integer. We find that $t_{\rm dip}$ increases as one approaches the flat band limit; this is consistent with Fig.\ \ref{fig2}(d), where $t_{\rm dip}^{-1}$ is plotted as a function of $\Lambda_F/(\hbar \Omega_1)$. $t_{\rm dip}$ decreases with increasing $\Lambda_F$, which itself increases with $w_0$ (as shown in the inset of Fig.\ \ref{fig2}(d)). The plateaus of $t_{\rm dip}^{-1}$ seen in Fig.\ \ref{fig2}(d) occurs due to the integer nature of $n_d$; a change in $n_d$ requires a finite change in $\Lambda_F$ and hence $w_0$.  Since thermalization occurs after the dip in the SFF, these results indicate the possibility of tuning to a large prethermal timescale by changing $w_0$; this tunability has no analogue for single-rate protocols. 

\begin{figure}
\rotatebox{0}{\includegraphics*[width= 0.48 \linewidth]{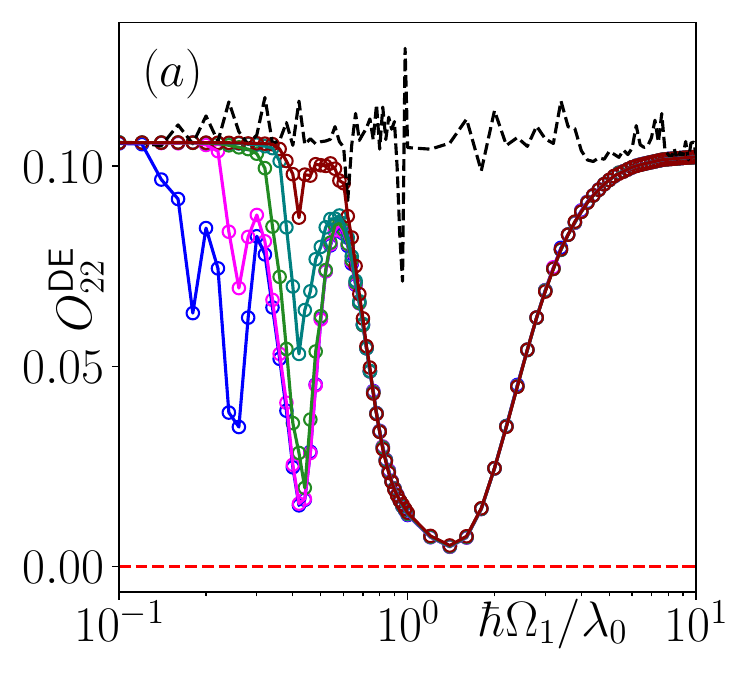}}
\rotatebox{0}{\includegraphics*[width= 0.48 \linewidth]{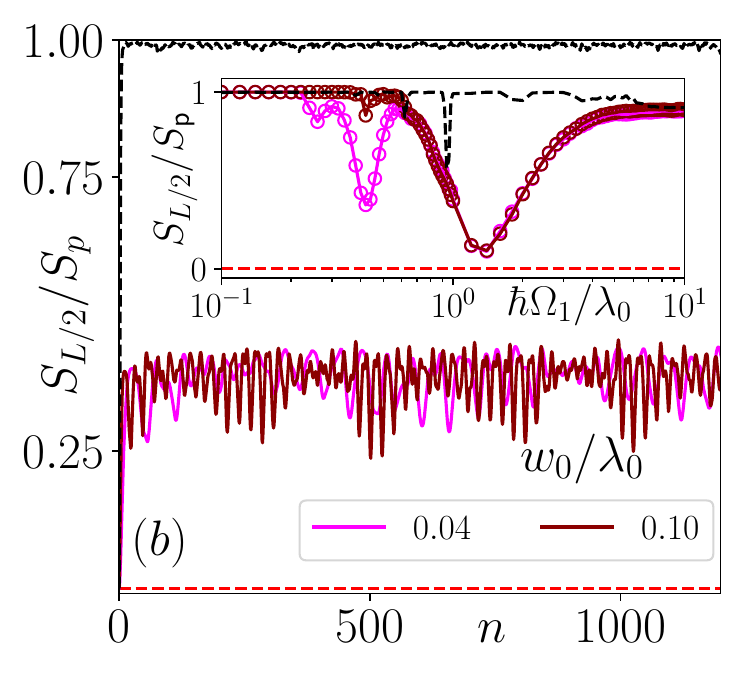}}
\rotatebox{0}{\includegraphics*[width= 0.48 \linewidth]{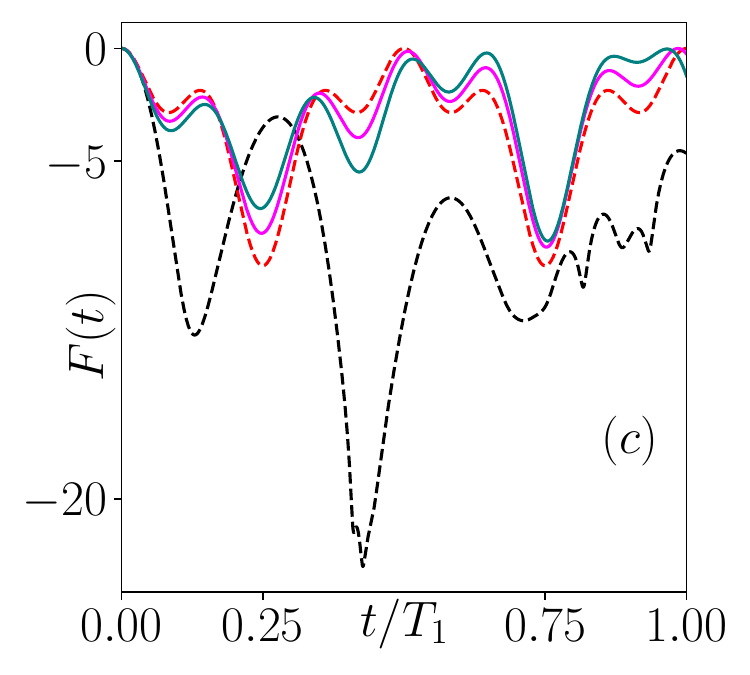}}
\rotatebox{0}{\includegraphics*[width= 0.48 \linewidth]{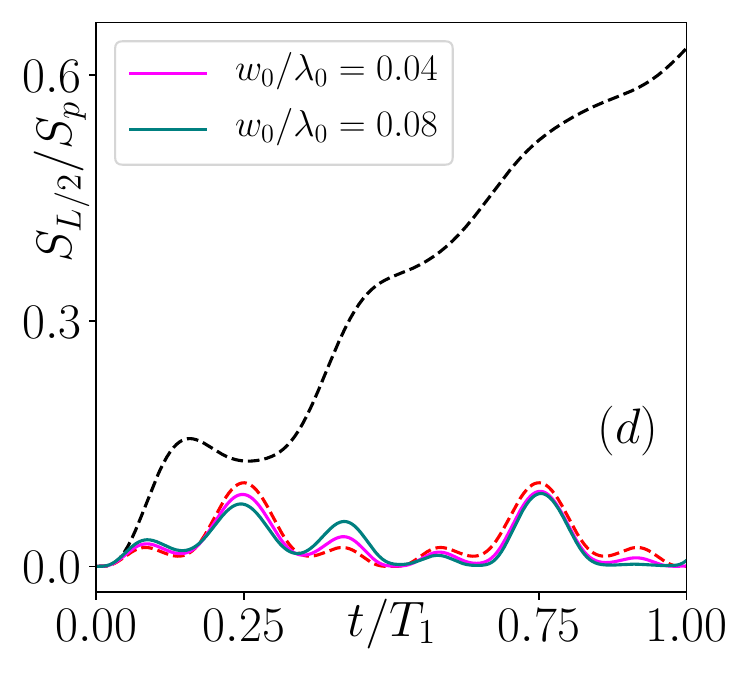}}
\caption{(a) Plot of $O_{22}^{\rm DE}$ as a function of $\hbar \Omega_1/\lambda_0$ for $L=26$ and several representative $w_0/\lambda_0$. The colors for different $w_0$ are same as in Fig.\ \ref{fig2}(a). (b) Plot of $S_{L/2}/S_p$ as a function of $n$ for $L=24$, $\hbar \Omega_1/\lambda_0=1$  and several $w_0/\lambda_0$. The inset shows a plot of long-time averaged value of  $S_{L/2}/S_p$ (averaged over $200$ cycles around $n=1900$) as a function of $\hbar \Omega_1/\lambda_0$ for several $w_0/\lambda_0$. For both plots $w_1/\lambda_0=1$. Plot of (c) $F(t)$ and (d) $S_{L/2}(t)/S_p$ as a function of $t/T_1$ ($0\le t/T_1\le 1$) for $w_1/\lambda_0=1$, $L=26$, and several $w_0/\lambda_0$ showing 
coherent revival of excitation for the two-rate protocol during the micromotion. The black (red) dashed line 
corresponds to $w_1/\lambda_0=0(1)$ and $w_0/\lambda_0=1(0)$. The drive used corresponds to the square-pulse protocol (Eq.\ \ref{sqp}) for (a) and (b) and cosine protocol (Eq.\ \ref{cosp}) for (c) and (d) with $\nu=3$. See text for details.  \label{fig3}}
\end{figure} 
  
{\it Correlation and Entanglement}: Typical local correlation functions for a finite Rydberg chain are known to reach the Floquet ETH predicted diagonal ensemble (DE) value \cite{flscar1}. In what follows, we study the DE value of the correlation function, $O_{j2}^{\rm DE} = {\rm Tr}[\rho_D \hat n_j \hat n_{j+2}]$ starting the from the vacuum (all spin down; $|0\rangle$) initial state. Here $\rho_D$ denotes the density matrix corresponding to the diagonal ensemble for the square-pulse protocol \cite{rev2}; $O_{j2}^{\rm DE}$ coincides with  $O_{j 2}(nT_1) = \langle \psi(nT_1)| \hat n_j \hat n_{j+2}|\psi(nT_1)\rangle$ for $n \to \infty$. The plot $O_{22}^{\rm DE}$ is shown in Fig.\ \ref{fig3}(a) as a function of $\Omega_1$ for $w_1/\lambda_0=1$, $K_0=0$, $P_0=1$, and $\nu=3$.  For the single-rate drive, $O_{j2}^{\rm DE} \simeq 0.105$ as shown in black dashed line in Fig.\ \ref{fig3}(a) \cite{flscar1}; in contrast, for $w_0=0$, $O_{22}$ remain pinned to its initial zero value for all $\Omega_1$ (red dashed line). In addition, for $w_0/\lambda_0 \le 0.1$ and $0.5 \le \hbar \Omega_1/\lambda_0 \le 5$, it exhibits a broad dip whose width is almost independent of $w_0$ and remains below the ETH predicted value. This behavior is qualitatively different from that obtained from single rate drive; an analytical understanding of this feature for small system sizes is presented in Ref.\ \cite{supp1}. 

A similar signature is found by studying the half-chain entanglement entropy $S_{L/2}$ of the driven chain \cite{flscar1,flfrag1}. We find that for $w_0/\lambda_0 \le 0.1$, $S_{L/2}$ never reaches its Page value $S_p$ \cite{pageref} for $n\le 10^3$ cycles signifying the absence of an infinite temperature steady state. Instead, $S_{L/2}/S_p$ oscillates around an average $\sim 0.27$ (Fig.\ \ref{fig3}(b)) for $\hbar \Omega_1/\lambda_0=1$, $K_0=0$, and $w_0/\lambda_0=0.04$ and $0.1$. This behavior is in sharp contrast to the single rate drive protocol where $S_{L/2}$ reaches $S_p$ within a few drive cycles (black dashed line in Fig.\ \ref{fig3}(b)). For $w_0=0$, $S_{L/2}=0$ as expected in the perfect flat band limit (red dashed line in Fig.\ \ref{fig3}(b)). The inset shows the  late-time behavior of $S_{L/2}/S_p$ obtained by averaging over $200$ cycles around $n=1900$; we find a similar dip as $O_{22}^{DE}$, showing lack of thermalization for all $w_0/\lambda_0\le 0.1$ and $0.5 \le \hbar \Omega_1/\lambda_0 \le 5$.  These results identify a wide range of $\Omega_1$ and $w_0$ around the flat band limit for which heating is significantly suppressed.

{\it Floquet Micromotion}: The micromotion corresponding to the two-rate drive protocol also exhibits a qualitatively different behavior compared to its single-rate counterpart. To understand this, we study the logarithm of the fidelity $F(t)= \ln |\langle \psi(0)|\psi(t)\rangle|^2$ and $S_{L/2}(t)$ of such a driven chain using the cosine drive protocol (Eq.\ \ref{cosp}) and starting from $|0\rangle$. A plot of $F(t)$ and $S_{L/2}(t)/S_p$, shown in Figs.\ \ref{fig3}(c) and (d) for $\hbar\Omega_1/\lambda_0=1=w_1/\lambda_0$ and several representative $w_0/\lambda_0$, brings out this difference. We find that $F(t)$ and $S_{L/2}(t)/S_p$ for the two-rate protocol shows clear signature of coherent reversal of excitations representing an almost perfect echo \cite{echoref1,echoref2} as can be seen from their non-monotonic nature (Fig.\ \ref{fig3}(c) and (d)); both of these quantities display oscillatory behavior close to their initial values for all $w_0 \le 0.1$. These features indicate that the state of the driven system almost comes back to itself at $t=T_1$; such an echo is exact in the flat-band limit ($w_0=0$). In contrast, the single-rate drive protocol leads to strong decay of $F(t)$ and fast growth of $S_{L/2}(t)/S_p$ consistent with fast spreading of the initial state leading to rapid heating. For the two-rate drive protocol, the micromotion displays a reflection symmetry around $t=T_1/2$ for $w_0=0$: $F(t)= F(T_1-t)$ and $S_{L/2}(t)= S_{L/2}(T_1-t)$ for all $t\le T_1$. This is an exact symmetry of the dynamics in the flat-band limit for both cosine and square-pulse protocols \cite{supp1}.

{\it Discussion}: The two-rate protocols studied in this work provide a way to realize {\it exact Floquet flat bands} for a large class of ergodic driven Hamiltonians. Such flat bands provide a starting point for studying ultra-strong correlation; our work provides the first Floquet version of this phenomenon. Here we concentrate on heating reduction due to the presence of such flat bands and the resultant coherent reversal of excitation formation in the micromotion of such systems. Such heating suppression is beneficial for quantum state preparation and qubit manipulation in driven systems. We note that similar two-rate protocols have already been implemented in experiments on Floquet-Hubbard models \cite{twoexp}. We expect our work to have several future extensions; for example, introduction of disorder through $w_0$ shall allow us to study many-body localization in such driven systems. It would also be interesting to examine these two-rate protocols to engineer time-translation-symmetry-breaking, thereby realizing prethermal discrete time crystals \cite{pdtc1}. We intend to address these and other related issues in future studies.

To conclude we have identified a class of two-rate drive protocols which leads to exact Floquet flat bands and hence to strong violation of ETH. We have identified a perturbative regime around such flat bands where heating is strongly suppressed. Our study provides an yet unexplored avenue for heating reduction in driven quantum systems and is expected to have applications in quantum state preparation~\cite{floqstp}, quantum simulation~\cite{floqsim}, and quantum metrology~\cite{floqmetro}.

\section{Acknowledgements} KS thanks DST, India for support through SERB project JCB/2021/000030 and Arnab Sen for discussions. SC thanks DST, India for support through SERB project SRG/2023/002730. TB thanks Mainak Pal for discussions.

\end{document}


\title{Supplementary Materials to Exact Floquet flat band and heating suppression via two-rate drive protocols}

\author{Tista Banerjee$^1$, Sayan Choudhury$^2$, and K. Sengupta$^1$}
\affiliation{$^1$School of Physical Sciences, Indian Association for the Cultivation of
Science, Jadavpur, Kolkata-700032, India. \\ $^2$Harish Chandra Research Institute, A CI of Homi Bhabha National Institute,
Chhatnag Road, Jhunsi, Prayagraj, Uttar Pradesh 211019, India. }

\date{\today}

\maketitle

\section{Additional numerical results} 

\begin{figure}
\rotatebox{0}{\includegraphics*[width= 0.98 \linewidth]{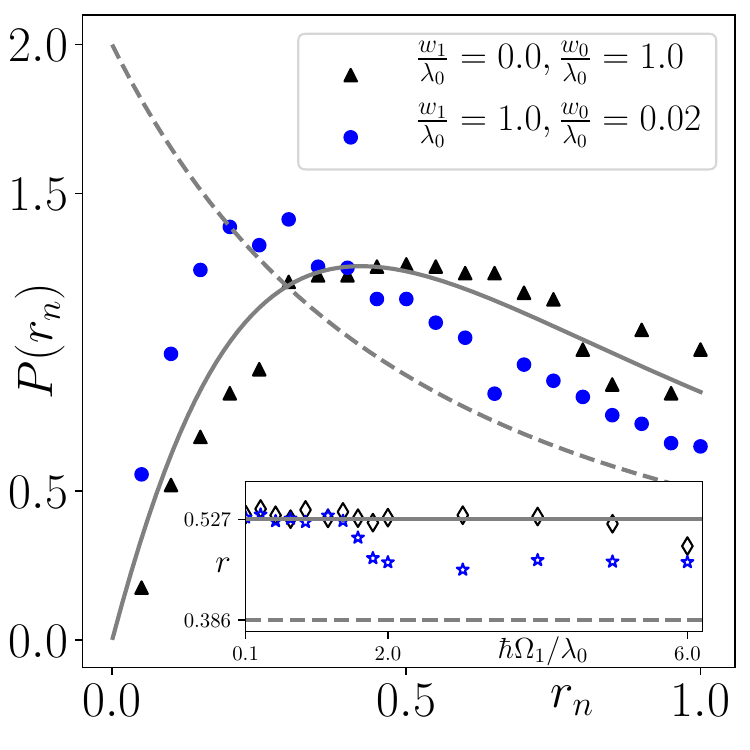}}
\caption{Plot of $P(r_n)$ as a function of $r_n$ for the single-rate drive with $w_0=1$ and $w_1=0$ (black triangles) and the two-rate drive  
for $w_0=0.02$ and $w_1=1$ (blue circles) for $\hbar \Omega_1/\lambda_0=3$. The grey solid line corresponds to COE distribution while the grey dotted line indicates Poissonian statistics. The inset shows a plot of $r$ as a function of $\hbar \Omega_1/\lambda_0$ for both protocols. For all plots $\nu=3$ and $L=26$ and we have considered eigenstates that belong to the zero total momentum ($K_0=0$) and even parity ($P_0=1$) sector. 
\label{figs1}}
\end{figure}

The numerical results used in the main text uses exact diagonalization (ED). To use this technique for periodic dynamics, we decompose the evolution operator $U(T_1,0)$ 
into finite number of Trotter steps. For the square pulse protocol with $\Omega_2/\Omega_1= \nu$, one needs $n_0=2\nu$ such steps. The Hamiltonian is diagonalized for each of these steps to 
obtain corresponding energy eigenvalues and  eigenstates. These, for the step $j$, are denoted by $E_{s_j}$ and $|s_j\rangle$ respectively. Then the evolution operator can be 
written in terms of these eigenstates and eigenvalues as 
\begin{eqnarray} 
U(T_1,0) &=& \prod_j \sum_{s_j} e^{-i E_{s_j} T_1/(2 \nu \hbar)} |s_j\rangle \langle s_j| \label{sqpuexp}
\end{eqnarray}
where $T_1=2\pi/\Omega_1$. A numerical diagonalization of $U(T_1,0)$ leads to its eigenstates $|p\rangle$ and eigenvalues $\alpha_p= e^{-i \theta_p(T_1)}$ (the form of the eigenvalues follows from unitary nature of $U$).
Using the relation $U(T_1,0)= \exp[-i H_F T_1/\hbar]$, one can read off the Floquet eigenvectors and eigenvalues as $|p\rangle$ and $E_p^F= \hbar \arccos[{\rm Re} (\alpha_p)]/T_1$ respectively. The folding of the 
Floquet eigenvalues to the first Floquet Brillouin zone is achieved by choosing $\arccos$ to yield a value within the first quadrant. For the cosine protocol an analogous procedure is carried out 
with the difference that the number of steps need to be large; in our numerics we find that choosing $n_0=500$ steps provide numerically accurate results which can be ascertained from the fact 
that increasing $n_0$ do not lead to significant change in values of $E_p^F$.

To supplement our results on the Floquet eigenstates in the main text, we first provide an analysis of the 
spectral statistics. To this end, we study the distribution $P(r_n)$ of the renormalized level-spacing ratio $r_n$ which is given, in terms of 
the Floquet level spacings $\epsilon_n^F= E_{n+1}^F - E_n^F$, by \cite{huse1,dalessio1,rev2}
\begin{eqnarray}
r_n &=& {\rm Min}[\epsilon_n^F, \epsilon_{n+1}^F]/ {\rm Max} [\epsilon_n^F, \epsilon_{n+1}^F] \label{lsr}
\end{eqnarray}
It is well-known that $P(r_n)$ obeys the circular orthogonal ensemble (COE) at low frequency with $r= \langle r_n\rangle = \sum_{r_n} P(r_n) r_n \simeq 0.527$; for the present system, at high frequency where the effective Floquet Hamiltonian 
$H_F \equiv H_{\rm PXP} =w_0 \sum_j \tilde \sigma_{j}^x$, $r \sim 0.536$ and $P(r_n)$ is well approximated by a Gaussian unitary ensemble (GUE). In the latter regime, the Floquet eigenvalues always remain within the first Floquet Brillouin zone (FBZ) and no unfolding is necessary. In contrast, in the former regime, majority of the eigenvalues need to be folded back in the first FBZ. In both these case, the states exhibits level repulsion leading to the respective ETH predicted value of $r$. However, for intermediate drive frequencies, only a finite fraction of the eigenvalues need to be folded back while the other stay within the first FBZ. Since the states which requires folding do not repel the unfolded ones \cite{santos1}, $r$ deviates from its GUE or COE values and exhibits a dip at intermediate frequencies.     

Fig.\ \ref{figs1} where the black triangles (blue circles) correspond to the single-rate (two-rate) drive protocol with $w_1=0$ and $w_0=1$ 
($w_1=1$ and $w_0=0.02$) and $\hbar \Omega_1=\lambda_0 =1$, confirms this expectation. The plot of $P(r_n)$ shows the expected COE feature for both drive protocols. However, for the two-rate drive protocol the peak of the distribution shifts to a lower value indicating lower value of $r$. This is confirmed in the inset where $r$ is plotted as a function of $\hbar\Omega_1/\lambda_0$. We find a significant dip in $r$ in the intermediate frequency regime $2 \le \hbar \Omega_1/\lambda_0 \le 5$ for the two-rate protocol where the single rate value is close to its COE predicted value; this feature is consistent with 
lower bandwidth of the spectrum leading to persistence of partial folding of Floquet eigenstates till a much lower frequency.  

\begin{figure}
\rotatebox{0}{\includegraphics*[width= 0.98 \linewidth]{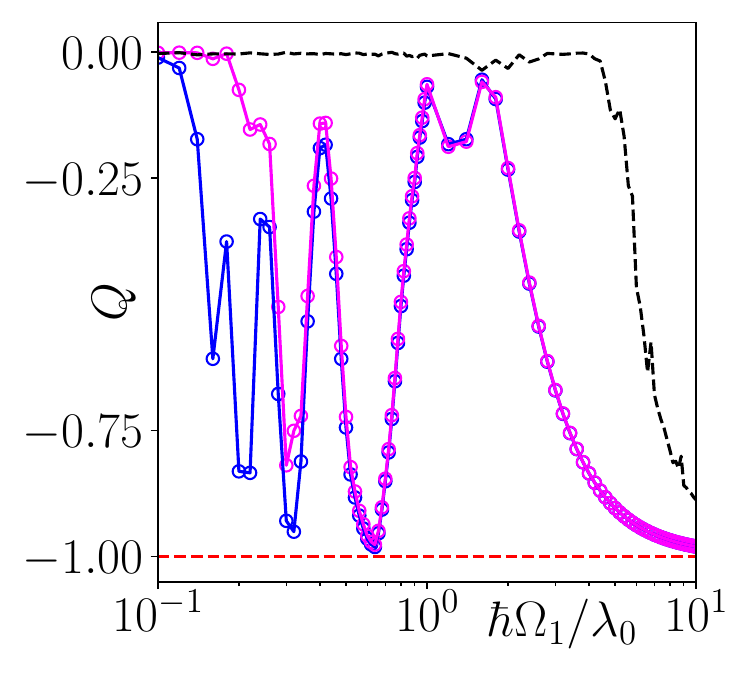}}
\caption{Plot of $Q$ as a function of $\hbar \Omega_1/\lambda_0$ for the single-rate drive with $w_0=1$ and $w_1=0$ (black dashed line) and the two-rate drive  for $w_1/\lambda_0=1$ and several representative $w_0/\lambda_0=0.04$ (pink circles) and $0.02$ (blue circles) and $0$ (red dashed line). For all curves $r=3$, $K_0=0$, and $L=22$. 
\label{figs2}}
\end{figure}

The reduction of heating for such two-rate drive protocols can be further understood from the dimensionless absorbed energy $Q$ defined as \cite{dalessio1} 
\begin{eqnarray} 
Q &=& \frac{\langle \psi(nT)|H_F^{{\rm av}} |\psi(nT)\rangle_{n \to \infty} - \langle H_F^{{\rm av}} \rangle_{\infty}}{ 
\langle H_F^{{\rm av}} \rangle_{\infty}-\langle \psi(0)|H_F^{{\rm av}} |\psi(0)\rangle} \label{qdef}
\end{eqnarray} 
where, $\langle .. \rangle_{\infty}$ denotes infinite-temperature average, $H_F^{{\rm av}} = \int_0^{T_1} H(t) dt/T_1 $ is the average Hamiltonian 
which coincides with the exact $H_F$ in the infinite frequency limit and is given by $H_F^{{\rm av}}= w_0 \sum_j \tilde \sigma_j^x$. Here we start from a superposition state 
$|\psi(0)\rangle =\sum_n c_n |n\rangle$ where $|n\rangle$ denotes Fock states with $n\le 3$ up-spins; $c_n$s are chosen so that the $|\psi(0)\rangle$ corresponds to equal superposition of ten 
lowest bit representations (in the Fock basis starting from all spin-down state) of the $k_{\rm total}=0$ momentum sector. This ensures that $\langle \psi(0)|H_F^{{\rm av}} |\psi(0)\rangle \ne 0$. 

It is clear that the limiting values of $Q$ are  $-1$ and $0$; these two limits correspond to its initial and infinite temperature values respectively. The plot of $Q$ as a function of $\hbar \Omega_1/\lambda_0$ is shown in Fig.\ \ref{figs2}; it clearly 
indicates that heating of the system is suppressed for a finite chain in the presence of the two-rate protocol for 
all $\hbar \Omega_1/\lambda_0 \ge 0.5$ and $w_0 \le 0.04$. We have checked that qualitatively similar behavior occurs for all $w_0 \le 0.1$ . In this regime $Q$ always remain negative indicating that the system does not heat up to infinite temperature; this is in sharp contrast to the single frequency drive as shown in Fig.\ \ref{figs2}. We have checked that the rise in $Q$ at large and intermediate drive frequencies is always slower for the two-rate protocol; in contrast, the nature and the position of the successive dips depend on the initial state chosen and the system size.

\begin{figure}
\rotatebox{0}{\includegraphics*[width= 0.98 \linewidth]{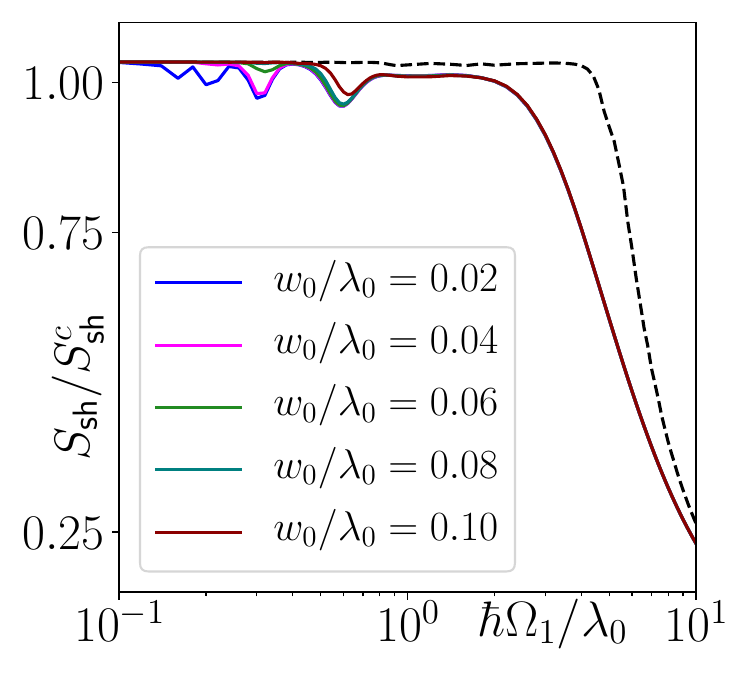}}
\caption{Plot of $S_{\rm sh}/S_{\rm sh}^c$ as a function of $\hbar \Omega_1/\lambda_0$ for the single-rate drive with $w_0=1$ and $w_1=0$ (black dashed line) 
and the two-rate drive protocol with $w_1/\lambda_0=1$ and several representative $w_0/\lambda_0$. For all plots $\nu=3$ and $L=26$, $K_0=0$ and $P_0=1$. 
\label{figs3}}
\end{figure}

The lack of heating in the presence of the two-rate drive protocol occurs due to the restricted spreading of the driven wavefunction in the Hilbert space. This can also be corroborated by studying the Shannon entropy. The Shannon entropy is given, in terms of the overlap 
\begin{eqnarray}
c_p^{q_0} &=& \langle p|q_0\rangle \label{oveq} 
\end{eqnarray} 
of the exact Floquet eigenfunction $|p\rangle$  with those ($|q_0\rangle$) of $H_F^{{\rm av}}$, by 
\begin{eqnarray}
S_{\rm sh} &=& -\sum_{p,q_0=1..{\mathcal D}} |c_p^{q_0}|^2 \ln |c_p^{q_0}|^2/{\mathcal D}. \label{shen}
\end{eqnarray} 
It saturates to its  circular ensemble value $S_{\rm sh}^c \sim \ln 0.48 {\mathcal D}$ for an infinite temperature steady state and is zero when $H_F^{{\rm av}}$ coincides with $H_F$. 
A plot of $S_{\rm sh}/S_{\rm sh}^c$, shown in Fig.\ \ref{figs3}, as a function of frequency $\hbar \Omega_1/\lambda_0$ indicates a much lower value compared to the 
single rate drive for a wide range of $\Omega_1$ and for all $w_0/\lambda_0 \le 0.1$ indicating significantly lower spread in Hilbert space.

\begin{figure}
\rotatebox{0}{\includegraphics*[width= 0.98 \linewidth]{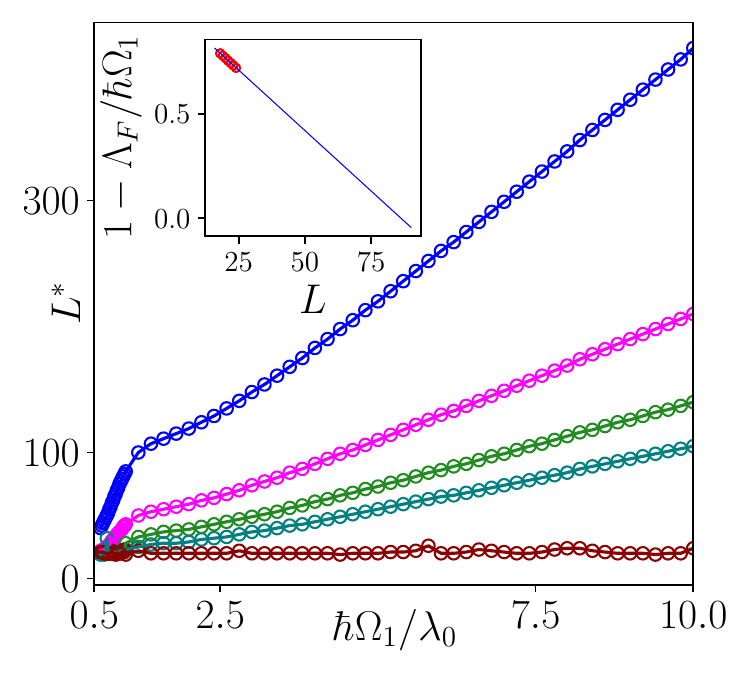}}
\caption{Plot of $L^{\ast}$ as a function of $\hbar \Omega_1/\lambda_0$ for several $w_0/\lambda_0$ and $w_1/\lambda_0=1$. The inset shows a plot of $1-\Lambda_F/(\hbar \Omega_1)$ as a function of $L$ for $w_0/\lambda_0=0.02$; $L^{\ast}$ is obtained from extrapolation of such plots such that  $\Lambda_F/(\hbar \Omega_1)=1$ for $L=L^{\ast}$. The color coding for different $w_0$ are same as in Fig.\ \ref{figs3}.
\label{figs4}}
\end{figure}

Next, we study the approach of the driven finite-sized chains to their thermodynamic limit. We note that the {\it perfect flat band limit} corresponds to a violation of ETH even in the thermodynamic limit. However, away from such fine tuned points, {\it i.e.}, for any finite $w_0$, such driven systems are expected to ultimately converge to the ETH predicted infinite temperature results as $L \to \infty$. To verify this and to understand how fast the system approach this limit, we carry out a extrapolation of $\Lambda_F/(\hbar \Omega_1)$ as shown in the inset of Fig.\ \ref{figs4} for $\hbar \Omega_1/\lambda_0=w_1/\lambda_0=1$ and $w_0/\lambda_0=0.1$. This allows us to extract a system size $L^{\ast}$ beyond which $\Lambda_F \sim \hbar \Omega_1$ which is the ETH predicted value. A plot of $L^{\ast}$ as a function of $\hbar \Omega_1/\lambda_0$ is shown in Fig.\ \ref{figs4}; we find that $L^{\ast}$ is generically larger for the two-rate drive protocol at any $\Omega_1$. Moreover, it increases with decreasing $w_0$; these features allow one to identify a parameter regime where a clear reduction of heating can be achieved for finite size systems.

\begin{figure}
\rotatebox{0}{\includegraphics*[width= 0.98 \linewidth]{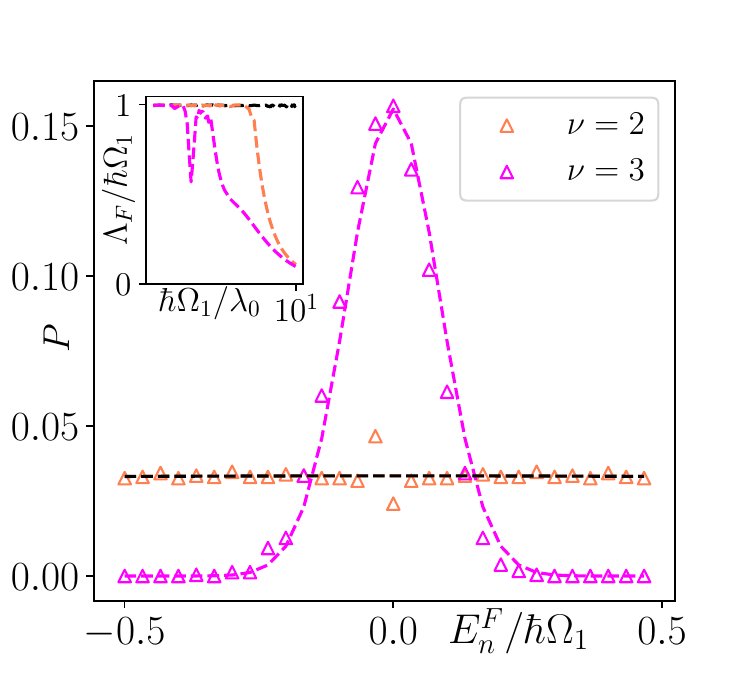}}
\caption{Plot of the distribution $P$ for $r=2$ (orange triangles) and $r=3$ (pink triangles) with $w_0/\lambda_0=0.04$, $L=24$, $K_0=0$, $P_0=1$, $w_1/\lambda_0=1$, and $\hbar \Omega_1/\lambda_0=1$. The black dashed line correspond to a single rate protocol with $w_0/\lambda_0=1$ and $w_1=0$. The eigenvalues were obtained by decomposing $U(T,0)$ into $N_0=500$ Trotter steps, followed by their diagonalization. A further increase in Trotter steps did not lead to significant change in the obtained eigenvalues. The inset shows a plot of $\Lambda_F/(\hbar \Omega_1)$ as a function of $\hbar \Omega_1/\lambda_0$ for $L=22$, $K_0=0$, and $P_0=1$. See text for details. 
\label{figs6}}
\end{figure}

Finally, we present exact numerical results for the distribution of the Floquet eigenstate, $P(E_n/\hbar \Omega_1) \equiv P$ for cosine protocol with $r=2,3$, $w_0/\lambda_0=0.04$, $w_1/\lambda_0=1$, and $\hbar \Omega_1/\lambda_0=1$ (see Eq. 2 of the main text) in Fig.\ \ref{figs6}. The corresponding inset shows a plot of the normalized Floquet bandwidth $\Lambda_F/(\hbar \Omega_1)$ as a function of the drive frequency $\hbar \Omega_1/\lambda_0$. The plots of $P$ for $r=3$ are qualitatively similar to their square-pulse counterpart discussed in the main text. In contrast, for $r=2$, where $w_0=0$ does not lead to a flat band, $P$ is almost flat; moreover $\Lambda_F$ converges to the expected value $\hbar \Omega_1$ quite fast. These features are qualitatively similar to the single drive protocol. These differences show the qualitative distinction between the dynamics obtained for odd integer values of $r$ where $w_0=0$ yield exact flat bands with their even integer counterparts.

\section{Floquet Hamiltonian at large drive amplitude} 

In this section, we are going to present analytic, albeit perturbative, results for the Floquet Hamiltonian for the case where amplitude of one of 
the drives is large: $\lambda_0 \gg w_0, w_1$. In this regime, once can use the standard Floquet perturbation theory (FPT) to obtain the Floquet Hamiltonian \cite{revfpt,dsen1,tb1}. In what follows, we shall consider a square-pulse protocol for $r=3$ (Eq. 1 of the main text). 

In the large drive amplitude regime, the Hamiltonian can be written as $H= H_0(t) + H_1(t)$ where 
\begin{eqnarray}
H_0(t) &=& \lambda(t) \sum_j \sigma_j^z \label{hamdef0} \\
H_1(t) &=& H_1^+ + H_1^-, \quad H_1^{\pm} =  (w_0 +w_1(t)) \sum_j \tilde \sigma_j^{\pm} \nonumber
\end{eqnarray} 
Within the FPT, the zeroth order Floquet Hamiltonian is given by 
\begin{eqnarray} 
U_0(t,0) &=& e^{-i \lambda_0 t \sum_j \sigma_j^z/\hbar} \theta(T_1/2-t) \nonumber\\
&& + e^{-i \lambda_0 (T_1-t) \sum_j \sigma_j^z/\hbar} \theta(t-T_1/2) \label{uoeq}
\end{eqnarray}
Since $U_0(T_1,0)=I$, this leads to $H_F^{(0)}=0$. To obtain the first-order term, we first define Fock states $|m\rangle$ with $m$ additional up-spins; note that $|m\rangle$ corresponds to multiple Fock states since they may have different positions of these spins. To compute the first order 
Floquet Hamiltonian, we consider the matrix element of $U_1(T_1,0)$ between these states. We first note that 
\begin{eqnarray}
U_1(T_1,0) &=& \sum_{s=\pm} \left(\frac{-i}{\hbar}\right) \int_0^{T_1} dt U_0^{\dagger}(t,0) H_1^s (t) U_0(t,0) \label{u1def}
\end{eqnarray} 
and define integrals 
\begin{eqnarray} 
I_1 (t_1,t_2,s) &=& \int_{t_1}^{t_2} dt' e^{-2 i \lambda_0 s t'/\hbar}  \nonumber\\
I_2 (t_1,t_2,s) &=& \int_{t_1}^{t_2} dt' e^{-2 i \lambda_0 s (T_1-t')/\hbar}  \label{intdef1}
\end{eqnarray} 
where $s=\pm 1$. Using Eqs.\ \ref{u1def} and \ref{intdef1}, a few lines of straightforward algebra yields
\begin{eqnarray} 
\langle m|U_1(T_1,0)|n\rangle &=& \left(\frac{-i}{\hbar}\right) \sum_{s=\pm 1} \delta_{m, n+s} c_s^{(1)} \label{u1sol1} \\
c_s^{(1)} &=&  \int_0^{T_1/2} dt' e^{-2 i \lambda_0 s t'/\hbar} (w_0 + w_1(t)) \nonumber\\
&& + \int_{T_1/2}^{T_1} dt' e^{-2 i \lambda_0 s (T_1-t')/\hbar} (w_0 +w_1(t')) \nonumber
\end{eqnarray} 
For $r=3$, the evaluation of the integrals need to be done by further subdivision of the integrals into time intervals of $T_1/6$. A straightforward computation yields
\begin{widetext}
\begin{eqnarray}
c_s^{(1)} &=& (w_0-w_1) [I_1(0,T_1/6,s) + I_1(T_1/3,T_1/2,s) +I_2(2T_1/3, 5T_1/6, s)] \nonumber\\
&& +(w_0+w_1) [I_1(T_1/6,T_1/3,s) + I_2(T_1/2,2T_1/3,s) +I_2(5T_1/6,T_1,s)] \label{cseq1}
\end{eqnarray} 
\end{widetext}
These integrals are straightforward to evaluate and leads to the final expression for $U_1(T_1,0)$ 
to be 
\begin{eqnarray} 
U_1(T_1,0) &=& \left(\frac{-i T_1}{\hbar} \right) \frac{w_0 \sin(\lambda_0 T_1/2)}{\lambda_0 T_1/2} \sum_{j,s=\pm} \tilde \sigma_j^{s} e^{-i \lambda_0 T_1 s/2}  \nonumber\\
\label{u1finaleq} 
\end{eqnarray}
This yields the first order Floquet Hamiltonian $H_F^{(1)} = (i\hbar/T_1) U_1(T_1,0)$ to be  
\begin{eqnarray} 
H_F^{(1)} &=& \frac{w_0 \sin(\lambda_0 T_1/2)}{\lambda_0 T_1/2} \sum_{j,s=\pm} \tilde \sigma_j^{s} e^{-i \lambda_0 T_1 s/2}  \label{firstfl1}
\end{eqnarray}
The Floquet hamiltonian yields flat bands for $w_0=0$ for all $T_1$. In addition, as shown in Ref.\ \cite{bm1}, it shows approximate flat bands
at first order for $\lambda_0 T_1= 2 n \pi$ where $n \in Z$ similar to the single-rate protocol studied earlier. 

In the single rate protocol studied earlier, $H_F$ only had odd-order terms in $w_0$. This originated 
from the presence of the operator $C_0= \prod_j \sigma_j^z$ which satisfied $C_0 U C_0^{-1} = U^{-1}$ leading to the requirement $\{H_F, C_0\}=0$. This necessitated that $H_F$ have only terms which are product of odd numbers of $\sigma_j^{\pm}$; thus only odd order terms (of $w_0$) contributed to 
$H_F$. However for the present case  when both $w_0$ and $w_1$ are finite, it is  easy to check that $C_0 U C_0^{-1} \ne U^{-1}$. This leads to an asymmetry of the spectrum around $E_F=0$ and  also allows for the second-order terms in $H_F$ which we now proceed to compute. 

To this end, we first compute $U_2(T_1,0)$ given by 
\begin{eqnarray} 
U_2(T_1,0) &=& \sum_{s_1,s_2=\pm } \left(\frac{-i}{\hbar}\right)^2 \int_0^{T_1} dt_1  U_0^{\dagger}(t_1,0) H_1^{s_1}(t_1) \label{u2eq1} \\ 
&& \times U_0(t_1,0)  \int_0^{t_1} dt_2 U_0^{\dagger}(t_2,0) H_1^{s_2}(t_2) U_0(t_2,0).    \nonumber
\end{eqnarray} 
The matrix elements of $U_2$ between the states $|m\rangle$ and $|n\rangle$ can be similarly computed and yields 
\begin{widetext} 
\begin{eqnarray} 
\langle m|U_2(T_1,0)|n\rangle &=& \left(\frac{-i}{\hbar}\right)^2  \sum_{s_1,s_2=\pm 1} c_{s_1,s_2}^{(2)} \delta_{m,n+s_1+s_2} \nonumber\\
c_{s_1 s_2}^{(2)} &=& A_1(T_1/2,s_1,s_2) + A_2(T_1, s_1, s_2) + A_3(T_1,s_1) A_4(T_1/2,s_2) \nonumber\\
A_1(T_1/2,s_1,s_2) &=& \int_0^{T_1/2} dt_1  e^{ -2 i\lambda_0 t_1 s_1/\hbar} (w_0+w_1(t_1)) \int_0^{t_1} dt_2 e^{ -2i\lambda_0 t_2 s_2/\hbar} (w_0+w_1(t_2)) \nonumber\\ 
A_2(T_1,s_1,s_2) &=& \int_{T_1/2}^{T_1} dt_1  e^{- 2 i\lambda_0 (T_1-t_1) s_1/\hbar} (w_0+w_1(t_1)) \int_{T_1/2}^{t_1} dt_2 e^{-2 i\lambda_0 (T_1-t_2) s_2/\hbar} (w_0+w_1(t_2)) \label{u2mat} \\
A_3(T_1,s_1) &=& \int_{T_1/2}^{T_1} dt_1  e^{-2 i\lambda_0 (T_1-t_1) s_1/\hbar} (w_0+w_1(t_1)) \quad
A_4(T_1/2,s_2) = \int_0^{T_1/2} dt_1  e^{- 2 i\lambda_0 t_1 s_2/\hbar} (w_0+w_1(t_1)) \nonumber 
\end{eqnarray} 
\end{widetext} 
To evaluate these integrals, we first consider $A_1$. To do this one needs to evaluate $w_1(t_1)$ and $w_1(t_2)$ at each time step. This leads to the 
result
\begin{widetext}
\begin{eqnarray} 
A_1(T_1/2,s_1,s_2) &=& (w_0-w_1)^2 (L_1 + L_4 + L_6) + (w_0^2-w_1^2)(L_2+L_5) + (w_0+w_1)^2 L_3 \nonumber\\
A_2(T_1,s_1,s_2) &=& (w_0+w_1)^2 (L'_1 + L'_4 + L'_6) + (w_0^2-w_1^2)(L'_2+L'_5) + (w_0-w_1)^2 L'_3 \nonumber\\
A_3(T_1,s_1) &=& (w_0-w_1) I_2(5T_1/6,2T_1/3,s_1) + (w_0+w_1) [I_2(T_1/2,5T_1/6,s_1) + I_2(2T_1/3, T_1,s_1)] \nonumber\\
A_4(T_1/2,s_2) &=& (w_0+w_1) I_1(T_1/6,T_1/3,s_2) + (w_0-w_1) [I_1(0,T_1/6,s_2) + I_1(T_1/3, T_1/2,s_2)] \label{secondintexp} 
\end{eqnarray} 
\end{widetext} 
The expressions for $A_3$ and $A_4$ in Eq.\ \ref{secondintexp} can be obtained in terms of the integrals $I_1$ and $I_2$ defined in Eq.\ \ref{intdef1}. In contrast, $A_1$ and $A_2$ requires evaluation of integrals $L_1..L_6$ and $L'_1..L'_6$ which are given by
\begin{widetext}
\begin{eqnarray} 
L_1 &=& \int_0^{T_1/2} dt_1 e^{-2 i \lambda_0 s_1 t_1/\hbar} \int_0^{t_1} dt_2 e^{-2 i \lambda_0 s_2 t_2/\hbar},
\quad L'_1 = \int_{T_1/2}^{T_1} dt_1 e^{-2 i \lambda_0 s_1 (T_1-t_1)/\hbar} \int_{T_1/2}^{t_1} dt_2 e^{-2 i \lambda_0 s_2 (T_1-t_2)/\hbar} \nonumber\\
L_2 &=& \int_{T_1/6}^{T_1/2} dt_1 e^{-2 i \lambda_0 s_1 t_1/\hbar} \int_0^{T_1/6} dt_2 e^{-2 i \lambda_0 s_2 t_2/\hbar},
\quad L'_2 = \int_{2 T_1/3}^{T_1} dt_1 e^{-2 i \lambda_0 s_1 (T_1-t_1)/\hbar} \int_{T_1/2}^{2T_1/3} dt_2 e^{-2 i \lambda_0 s_2 (T_1-t_2)/\hbar} \nonumber\\
L_3 &=& \int_{T_1/6}^{T_1/2} dt_1 e^{-2 i \lambda_0 s_1 t_1/\hbar} \int_{T_1/6}^{t_1} dt_2 e^{-2 i \lambda_0 s_2 t_2/\hbar},
\quad L'_3 = \int_{2T_1/3}^{T_1} dt_1 e^{-2 i \lambda_0 s_1 (T_1-t_1)/\hbar} \int_{2T_1/3}^{t_1} dt_2 e^{-2 i \lambda_0 s_2 (T_1-t_2)/\hbar} \nonumber\\
L_4 &=& \int_{T_1/3}^{T_1/2} dt_1 e^{-2 i \lambda_0 s_1 t_1/\hbar} \int_0^{T_1/6} dt_2 e^{-2 i \lambda_0 s_2 t_2/\hbar},
\quad L'_4 = \int_{5 T_1/6}^{T_1} dt_1 e^{-2 i \lambda_0 s_1 (T_1-t_1)/\hbar} \int_{T_1/2}^{2 T_1/3} dt_2 e^{-2 i \lambda_0 s_2 (T_1-t_2)/\hbar} \nonumber\\
L_5 &=& \int_{T_1/3}^{T_1/2} dt_1 e^{-2 i \lambda_0 s_1 t_1/\hbar} \int_{T_1/6}^{T_1/3} dt_2 e^{-2 i \lambda_0 s_2 t_2/\hbar},
\quad L'_5 = \int_{5 T_1/6}^{T_1} dt_1 e^{-2 i \lambda_0 s_1 (T_1-t_1)/\hbar} \int_{2 T_1/3}^{5 T_1/6} dt_2 e^{-2 i \lambda_0 s_2 (T_1-t_2)/\hbar} \nonumber\\
L_6 &=& \int_{T_1/3}^{T_1/2} dt_1 e^{-2 i \lambda_0 s_1 t_1/\hbar} \int_{T_1/3}^{t_1} dt_2 e^{-2 i \lambda_0 s_2 t_2/\hbar},
\quad L'_6 = \int_{5 T_1/6}^{T_1} dt_1 e^{-2 i \lambda_0 s_1 (T_1-t_1)/\hbar} \int_{5 T_1/6}^{t_1} dt_2 e^{-2 i \lambda_0 s_2 (T_1-t_2)/\hbar} \label{intdef2}
\end{eqnarray} 
\end{widetext} 
The evaluation of these integrals are tedious but straightforward. These lead to the expressions of $c_{s_1, s_2}^{(2)}$ and $U_2(T_1,0)$ can be expressed as
\begin{eqnarray}
U_2(T_1,0) &=& \left(\frac{-i}{\hbar}\right)^2 \sum_{j_1,j_2} \sum_{s_1,s_2=\pm 1} c_{s_1,s_2}^{(2)}  \sigma_{j_1}^{s_1} \sigma_{j_2}^{s_2} \nonumber\\
H_F^{(2)} &=& \left(\frac{i \hbar}{T_1} \right) (U_2(T_1,0) - U_1(T_1,0)^2/2) \label{hf2exp}
\end{eqnarray} 
where $U_1(T_1,0)$ is given by Eq.\ \ref{u1finaleq}.

While evaluating $H_F^{(2)}$ using  Eq.\ \ref{hf2exp} we find that it vanishes if $s_1=s_2$. Moreover it also vanishes if $j_1\ne j_2$ or $j_1\ne j_2\pm 1$. The last property ensures locality of the Floquet Hamiltonian. Furthermore we note that terms such as $\tilde \sigma_j^- \sigma_{j'}^+$ is identically zero for $j \ne j'$ due to the constraint on the spins. Finally, we find that $H_F^{(2)}=0$ if either $w_0=0$ or $w_1=0$; this ensures that our results are consistent with both the exact Floquet flat band limit ($w_0=0$) and the single rate drive protocol ($w_1=0$) where $H_F^{(2)}$ is known to vanish. A straightforward but somewhat cumbersome algebra, using Eqs.\ \ref{secondintexp}, and \ref{intdef2} yields 
\begin{eqnarray}
H_F^{(2)} &=& \frac{2 w_0 w_1 {\mathcal C}}{\lambda_0}  \sum_{j} \left( \sigma_j^z + (\tilde \sigma_{j}^+ \tilde \sigma_{j+1}^- +{\rm h.c.}) \right) \label{hfsecond} \\
{\mathcal C} &=& 6 [2\sin (x/6) -2 \sin (x/3) +\sin (x/2)]/x -1, \nonumber
\end{eqnarray} 
where $x=2 \lambda_0 T_1/\hbar$. Note that in the high frequency limit where $x\to 0$, $C(x) \to 0$ which is consistent with standard high-frequency Magnus expansion.

\begin{figure}
\rotatebox{0}{\includegraphics*[width= 0.98 \linewidth]{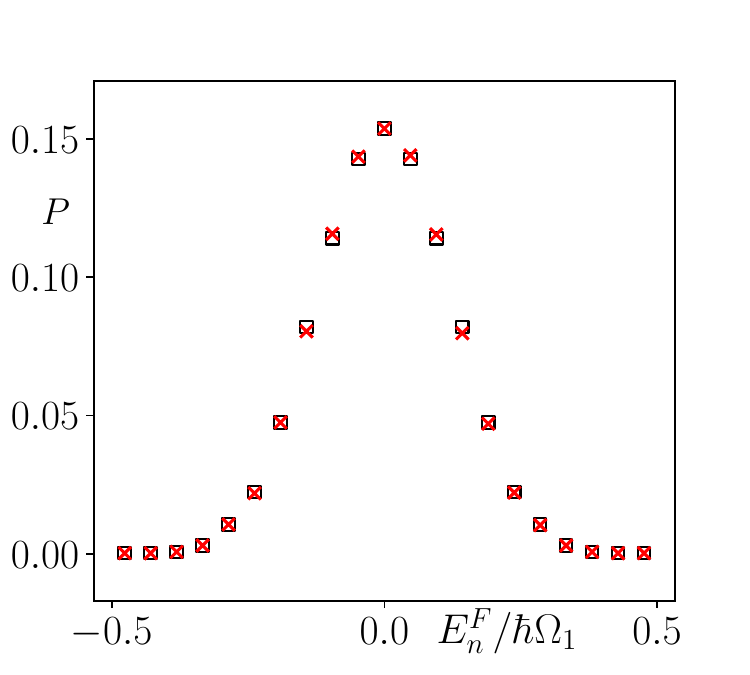}}
\caption{Plot of the eigenvalue distribution $P$ within the first FBZ for $ \hbar \Omega_1 / w_0 = 15$, $w_1 / w_0 = 0.1$, and $\lambda_0 / w_0 = 20$. The red crosses show results from second order FPT while the black open squares indicate those from exact numerics obtained using ED. For both plots $L=24$, $K_0=0$ and we have used square-pulse protocol with $\nu=3$. 
\label{figs7}}
\end{figure}

A comparison between the distribution of eigenstates, $P(E_n) \equiv P$, of the analytical Floquet Hamiltonian $H_F^{\rm an}= H_F^{(1)} + H_F^{(2)}$ and the exact Floquet Hamiltonian $H_F^{\rm exact}$ obtained using ED is shown in Fig.\ \ref{figs7} for $ \hbar \Omega_1 / w_0 = 15$, $w_1 / w_0 = 0.1$, and $\lambda_0 / w_0 = 20$. The result indicates excellent match between the exact numerics and the second-order FPT demonstrating that the Floquet eigenvalues are accurately captured in the large drive amplitude regime by the FPT. 

\section{Approximate freezing for $L=3$}

\begin{figure}
\rotatebox{0}{\includegraphics*[width= 0.98 \linewidth]{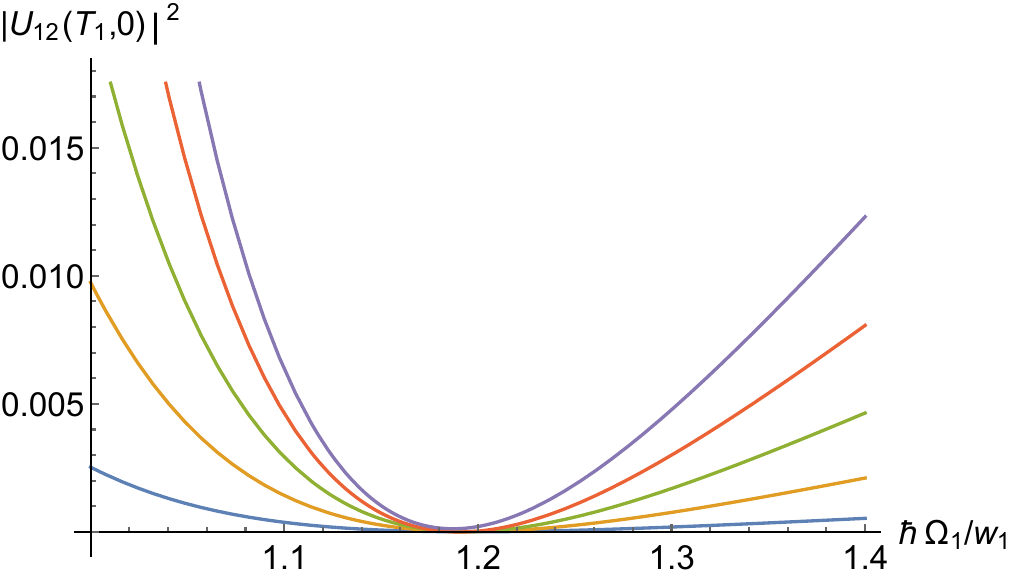}}
\caption{Plot of $|U_{12}(T_1,0)|^2$ as a function of $\hbar \Omega_1/w_1$ for $w_0=0.02 .. 0.1$ (bottom to top curves in steps of $0.02$) and $\lambda_0/w_1=1$. 
\label{figs5}}
\end{figure}

The phenomenon of dynamic freezing at special drive frequencies starting from the vacuum (all spin-down state) was studied for the single-rate drive protocol in Ref.\ \cite{bm2}. An analytical insight into the problem was obtained by studying the $L=3$ system analytically. Here we shall carry out a similar study for the two-rate protocol and aim to obtain an analytic understanding of the wide dip in $O_{22}^{\rm DE}$ discussed in the main text. To this end, we note that for the $L=3$ in the $k=0$ sector, there are just two states given by $|\phi_0\rangle= |\downarrow, \downarrow,\downarrow\rangle$ and $|\phi_1\rangle = (|\uparrow,\downarrow,\downarrow> + |\downarrow,\uparrow,\downarrow \rangle + |\downarrow,\downarrow,\uparrow\rangle)/\sqrt{3}$. In the space of these two states, one can define a $2 \times 2 $ matrix Hamiltonian given by
\begin{eqnarray} 
A(s_1, s_2) &=& s_1 \lambda_0 (1-\tau_z) + \sqrt{3} (w_0 + w_1 s_2)\tau_x  \label{adef} 
\end{eqnarray} 
where $s_{1,2} = \pm 1$ and all energies are measured from $E=-3\lambda_0$. The evolution operator for the square-pulse protocol given by Eq. (1) of the main text can then be written as 
\begin{widetext} 
\begin{eqnarray} 
U(T_1,0) &=& e^{-i A(-1,1) T_1/(6\hbar)}  e^{-i A(-1,-1) T_1/(6\hbar)} e^{-i A(-1,1) T_1/(6\hbar)}  e^{-i A(1,-1) T_1/(6\hbar)} e^{-i A(1,1) T_1/(6\hbar)} e^{-i A(1,-1) T_1/(6\hbar)} \label{uthsdef}  
\end{eqnarray} 
\end{widetext} 
The time evolution any local operator, starting from $|\phi_0\rangle$ is therefore controlled by $U_{12}(T_1,0)$ which we study as function of $\Omega_1$. For general $w_0$ and $w_1$, the expression of $U_{12}(T_1,0)$ is cumbersome. However, its expression for the case $w_0=w_1$ is particularly simple and is given by
\begin{widetext} 
\begin{eqnarray} 
U_{12}(T_1,0) &=& (2 i \sqrt{3} e^{-i x } \sin y )(z^2- 12 i \sin 2 x + 2 e^{-ix} \cos 2 x ((z^2+ 6 (e^{2 i x}-1)) \cos 2y - i z \lambda_0 \sin 2 y))/z^3
\end{eqnarray}
\end{widetext} 
where all quantities are scaled in units of $w_1$, $w_0=1$, $x=\lambda_0 T_1/(6\hbar)$, $z=\sqrt{12 + \lambda_0^2}$ and $y= z T_1/(6 \hbar)$. This leads to $U_{12}=0$ for $y= n_0 \pi$ where $n_0$ is an integer. These frequencies correspond to exact dynamic freezing for $w_0=w_1$; however such freezing is not stable against variation of $w_0$. For a generic $w_0 \ne w_1$, we do not find any exact dynamic freezing for $L=3$. In contrast, for $\lambda_0/w_1=1$ and $\hbar \Omega_1/w_1 \simeq 1.2$, we find a sharp dip in $U_{12}(T_1,0)$ which suggest a presence of approximate dynamic freezing. Remarkably, our numerical results for $w_0 \ne w_1$ suggest that this dip persists for $w_0 \le w_1$ at same $\Omega_1$; its breadth increases with decreasing $w_0$ and leads to a complete flat curve consistent with Floquet flat band for $w_0=0$. This feature is shown in Fig.\ \ref{figs5} where $|U_{12}(T_1,0)|^2$ is plotted as function of $\hbar \Omega_1/w_1$ for $\lambda_0/w_1=1$ and several representative $w_0$. For increasing $L$, we numerically find that such a dip remains leading slow evolution and the corresponding dip in $O_{22}^{\rm DE}$ and $S_E$ discussed in the main text. However, the position of the dip shifts to slightly higher frequency $\hbar \Omega_1/w_1 \simeq 1.4$. This is a consequence of contribution of additional evolution channels for $L>3$ and we have not been able to understand this analytically.  

\section{Symmetry of the micromotion}

In this section, we are going to provide a proof for a reflection symmetry of the drive protocol around $t=T_1/2$ 
for $w_0=0$. We shall provide an explicit proof using the square pulse protocol; the proof for a generic protocol 
follows the same line of reasoning. 

Consider a Hamiltonian 
\begin{eqnarray} 
H_1(t) &=& f_1(t) \hat O_1 + f_2(t) \hat O_2  \label{hdef} 
\end{eqnarray}
where ${\hat O}_{1(2)}$ are general operators. Let us also consider a square pulse protocol such $f_1(t)$ has a period $T_1$ and 
$f_2(t)$ has a period of $T_1/\nu$, where $\nu$ is an integer; the amplitudes of these drives are taken to be $f_{01}$ and $f_{02}$ respectively. 
We also define the Hamiltonian 
\begin{eqnarray} 
H(s_1,s_2) &=& f_{01} s_1 \hat O_1 + f_{02} s_2 \hat O_2 \label{hs1s2}
\end{eqnarray} 
where $s_{1(2)} = \pm 1$. For the square pulse protocol, the instantaneous Hamiltonian, $H_1(t)$, can always be written in terms of $H(s_1,s_2)$. 

In what follows, we shall consider micromotion under the action of $H(t)$. To this end, we first note that for $t\le T_1/(2 \nu)$, the 
evolution operator is given by
\begin{eqnarray} 
U_1(t,0) &=& e^{-i H(1,-1) t/\hbar} = \sum_{m^{+-}} e^{-i \epsilon_m^{+-} t/\hbar} |m^{+-}\rangle\langle m^{+-}| \nonumber\\
\label{u1exp} 
\end{eqnarray} 
where $\epsilon_m^{+-}$ denote eigenenergies corresponding $H(1,-1)$ and $|m^{+-}\rangle$ are the corresponding eigenstates. 
Next, let us consider the evolution operator between $T_1(1-1/(2\nu)) \le t\le T_1$. Defining $x_0= T_1/(2 \nu \hbar)$ and $\delta t= t -T_1(1-1/(2\nu))$, 
we find that the evolution operator can be written as 
\begin{widetext}
\begin{eqnarray}
U_2(t,0) &=& e^{-i H(-1,1) \delta t/\hbar  } e^{-i H(-1,-1) x_0} .... e^{-i H(-1,-1) x_0 } e^{-i H(1,1) x_0} 
.... e^{-i H(1,1) x_0 } e^{-i H(1,-1) x_0 } \nonumber\\
&=&e^{-i H(-1,1)(t-T_1(1-1/(2\nu)))/\hbar} e^{-i H(1,-1) T_1/(2 \nu \hbar)}   \label{u2exp1}
\end{eqnarray}
\end{widetext} 
where in the last line we have used the fact $H(s_1,s_2)=- H(-s1,-s2)$ so that the contributions from the intermediate exponential factors
cancel pairwise. Noting that $\epsilon_{m}^{s_1,s_2}= -\epsilon_{m}^{-s_1,-s_2}$ and 
$|m^{s_1,s_2}\rangle= |m^{-s_1,-s_2}\rangle$ (up to an unimportant global phase), we find 
\begin{eqnarray}
U_2(t,0) &=& \sum_{m^{+-}} e^{-i \epsilon_m^{+-} T_1/(2 \nu\hbar)} |m^{+-}\rangle\langle m^{+-}|  \nonumber \\
&& \times \sum_{m^{- +}} e^{-i \epsilon_m^{- +} (t-T_1(1-1/(2\nu)))/\hbar} |m^{-+}\rangle\langle m^{-+}| \nonumber\\
&=& \sum_{m^{+-}} e^{-i \epsilon_m^{+-} (T_1-t)/\hbar}  |m^{+-}\rangle\langle m^{+-}| \nonumber\\
&=& U_1(T_1-t,0)
\end{eqnarray} 
Thus, the evolution operator, for $t\le T_1/(2\nu)$, is symmetric around $T_1/2$. A similar derivation can be carried out 
for any $(m-1)T_1/(2\nu) \le t \le m T_1/(2\nu)$ and we find the symmetry
$U(t,0)= U(T_1-t,0)$ for all $t\le T_1/2$. Since $U$ controls the evolution of all operator correlations and expectations, we find that the dynamics 
is symmetric around $T_1/2$. The same relation can also be derived for other protocols using a suitable Trotter decomposition of $U$.